\documentclass[conference]{IEEEtran}
\ifCLASSINFOpdf
\else
\fi
\IEEEoverridecommandlockouts
\usepackage{cite}
\usepackage{amsmath,amssymb,amsfonts}
\usepackage{algorithmic}
\usepackage{fancybox}
\usepackage{pdfpages}
\usepackage{comment}
\usepackage{graphicx}
\usepackage{mathtools, cuted}
\usepackage{lipsum, color}
\usepackage{comment}
\DeclareUnicodeCharacter{00A0}{~}
\usepackage{textcomp}
\usepackage{xcolor}
\def\BibTeX{{\rm B\kern-.05em{\sc i\kern-.025em b}\kern-.08em
    T\kern-.1667em\lower.7ex\hbox{E}\kern-.125emX}}

\begin{document}
%
\title{Advanced Biophysical Model to Capture Channel Variability for EQS Capacitive HBC}
%
%
%

\author{
         \IEEEauthorblockN{Arunashish Datta, Mayukh Nath, David Yang and~Shreyas~Sen,~\IEEEmembership{Senior~Member,~IEEE}}\\
         \IEEEauthorblockA{School of Electrical and Computer Engineering, Purdue University}
         
\thanks{The authors are with the School
of Electrical and Computer Engineering, Purdue University, West Lafayette
IN, 47907 USA.
Corresponding Author: Shreyas Sen, e-mail: shreyas@purdue.edu}}
%
%

\markboth{Journal of \LaTeX\ Class Files,~Vol.~14, No.~8, August~2015}%
{Shell \MakeLowercase{\textit{et al.}}: Bare Demo of IEEEtran.cls for IEEE Journals}
%



\maketitle

\begin{abstract}
Human Body Communication (HBC) has come up as a promising alternative to traditional radio frequency (RF) Wireless Body Area Network (WBAN) technologies. This is essentially due to HBC providing a \textit{broadband} communication channel with enhanced signal security in the physical layer due to lower radiation from the human body as compared to its RF counterparts. An in-depth understanding of the mechanism for the channel loss variability and associated biophysical model needs to be developed before EQS-HBC can be used more frequently in WBAN consumer and medical applications. Biophysical models characterizing the human body as a communication channel didn't exist in literature for a long time. Recent developments have shown models that capture the channel response for fixed transmitter and receiver positions on the human body. These biophysical models do not capture the variability in the HBC channel for varying positions of the devices with respect to the human body. In this study, we provide a detailed analysis of the change in path loss in a capacitive-HBC channel in the electro-quasistatic (EQS) domain. Causes of channel loss variability namely: inter-device coupling and effects of fringe fields due to body's shadowing effects are investigated. FEM based simulation results are used to analyze the channel response of human body for different positions and sizes of the device which are further verified using measurement results to validate the developed biophysical model. Using the bio-physical model, we develop a closed form equation for the path loss in a capacitive HBC channel which is then analyzed as a function of the geometric properties of the device and the position with respect to the human body which will help pave the path towards future EQS-HBC WBAN design. 
\end{abstract}


%
\IEEEpeerreviewmaketitle

\section{Introduction}

\IEEEPARstart{L}{ow} power wearable devices have traditionally used radio frequency (RF) wireless communication operating at narrowband frequencies like 400 MHz, 900 MHz and 2.4 GHz as well as ultra wide band 3-10 GHz frequency bands.
Recently, the use of human body as a channel to communicate with on body devices has come up as a promising alternative to RF based WBAN technologies. This has prompted further studies on the characteristics of the human body as a communication channel. Human Body Communication (HBC) \cite{Health_Monitoring , ISLPED, Magnetic, Safety_Study, SocialHBC} has been shown to be a low power and physically secure alternative to RF WBAN approaches. Physical layer security \cite{NSR, CICC_2020} can be achieved by containing the transmitted signal within the body which can be done by using low frequency of operation and hence having larger wavelength to limit the radiation from human body. Using HBC upto a bandwidth of about 1 MHz can result in \textit{low radiation} from the body. HBC can also be used for high bandwidth applications as a broadband channel at frequencies upto 30 MHz \cite{DATE, JSSC, CICC} using \textit{100X lower energy} than RF WBAN technologies.

\begin{figure}[h!]
\centering
\includegraphics[width=0.8\columnwidth]{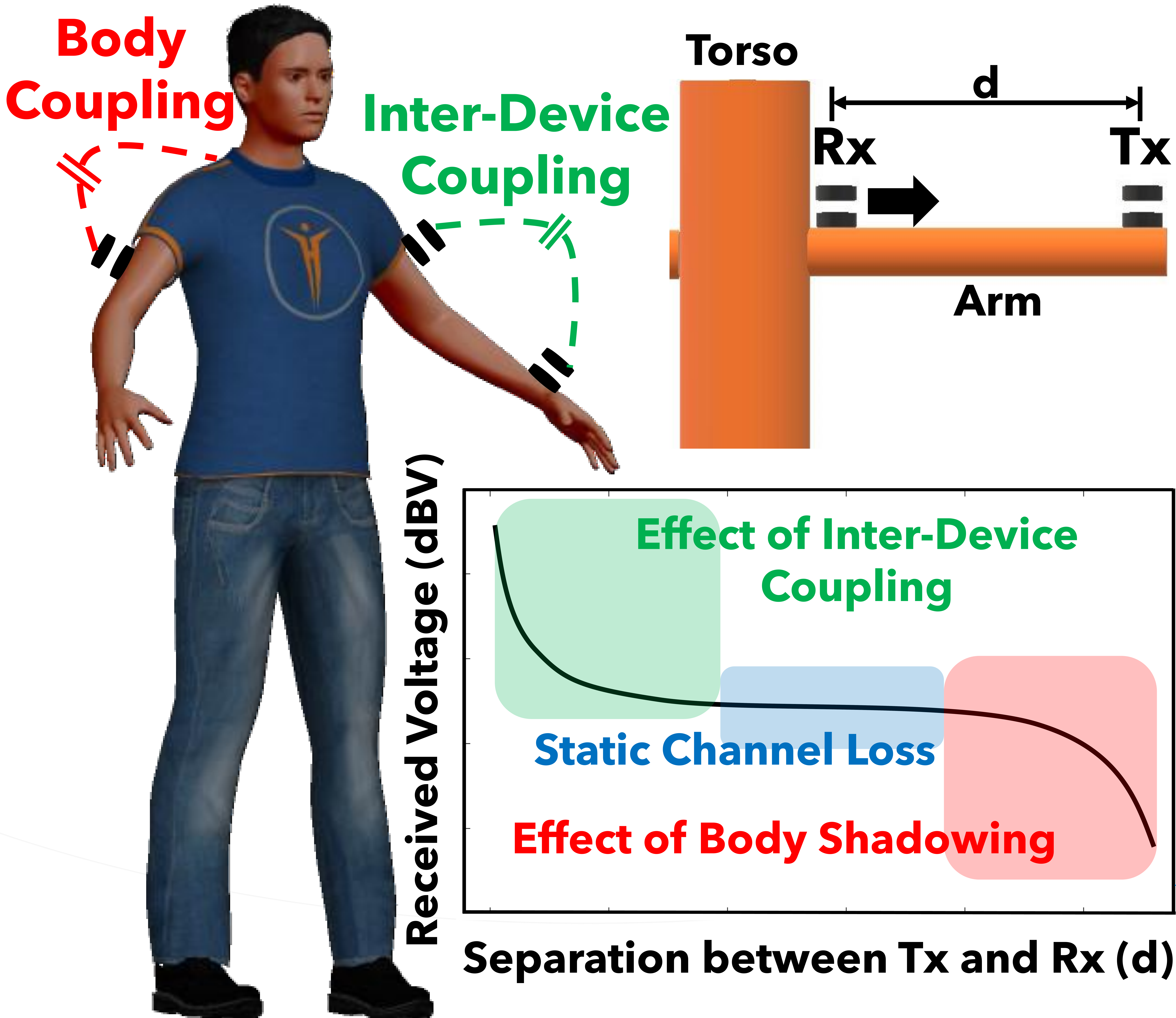}
\caption{Inter-device coupling and the shadowing effects of human body on the floating ground plate results in variation of channel loss with the position of the transmitting and receiving devices with respect to the human body. In this paper, we analyze how inter-device coupling and body shadowing affect the received voltage. FEM based simulations are used on a simplified human model shown in the figure to analyze the channel loss. Using this understanding, we develop a bio-physical circuit model which captures the channel loss variability for a capacitive HBC system in EQS domain.}
\label{fig:CG_CB_Model}
\end{figure}

\par
HBC can be categorized in two ways according to the method of operation, capacitive and galvanic HBC. Capacitive HBC \cite{Zimmerman} uses a single electrode to capacitively couple the signal to the body which makes up the forward path and parasitic capacitances from the device to the ground make up the return path for the circuit. In Galvanic HBC \cite{Wegmueller}, a pair of electrodes apply differential signal into the body where the fringe fields in the body medium are picked up by the receiver. Previous studies have analyzed the Capacitive EQS HBC channel for broadband low frequency applications. A thorough characterization of the human body as a communication channel is important in development of HBC as a viable method for WBAN applications in various domains. Biophysical models capturing the channel loss characteristics have been previously developed for capacitive HBC channels \cite{TBME19, 7463049, TBioCaS, Forward_PL}.

 
\begin{figure}[h!]
\centering
\includegraphics[width=0.8\columnwidth]{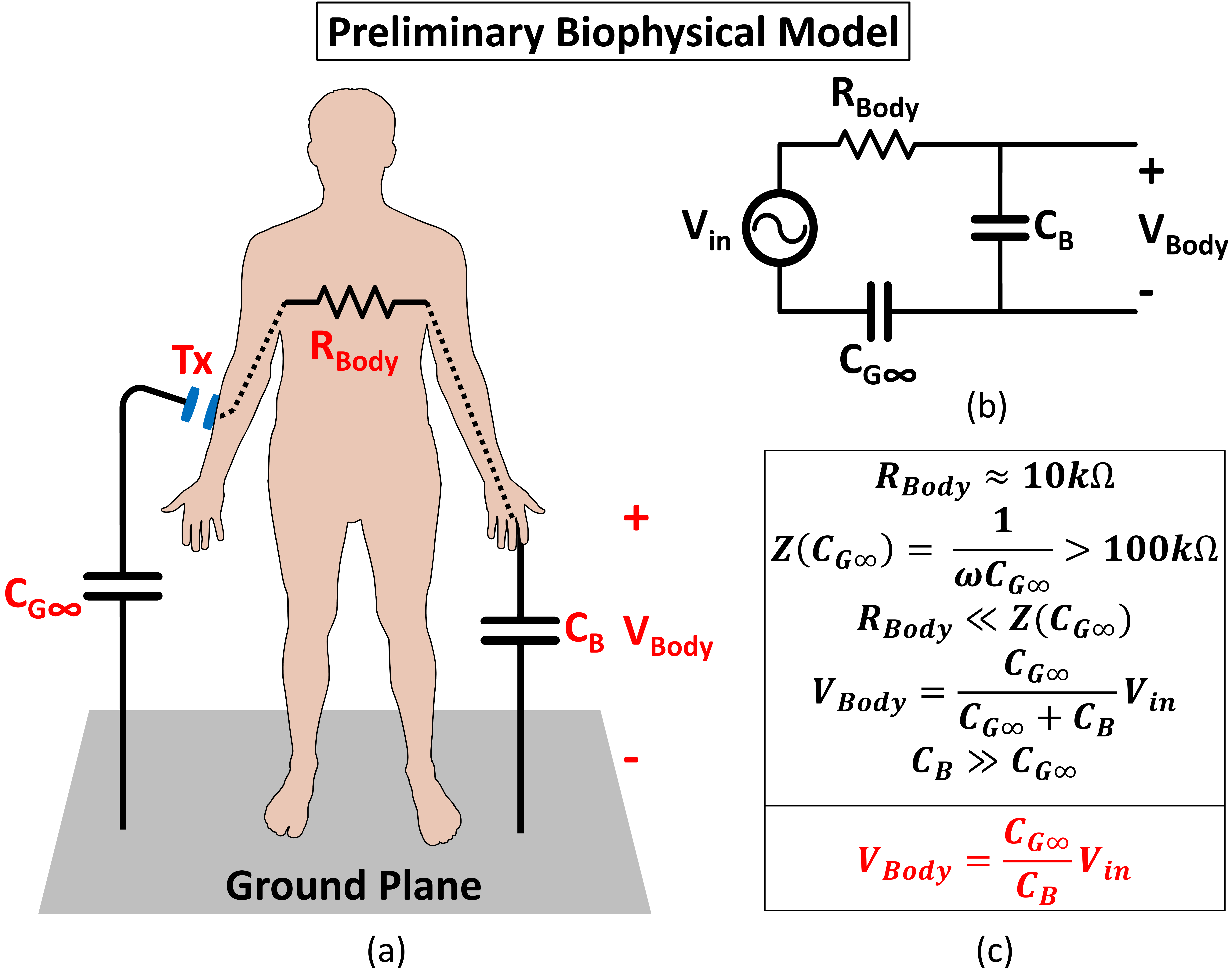}
\caption{(a) Simplified model of a Capacitive HBC system in EQS domain. (b) Circuit theoretical model of a capacitive HBC system. (c) Channel loss as a function of the return path capacitance $(C_G\infty)$ and body capacitance $(C_B)$.}
\label{fig:CG_CB_Circuit}
\end{figure}

However, these preliminary biophysical models consider the HBC channel to have a constant channel loss with respect to position of the transmitting and receiving devices and hence doesn't explore the variability in the channel loss. Our focus in this paper will be to explore the cause for channel variability in a capacitive EQS HBC channel and to develop the biophysical model to capture this variability. Fig. \ref{fig:CG_CB_Model} illustrates how in a real world capacitive HBC channel, the position of the transmitter and receiver affects the channel quality. In-depth analysis shows positional variance in the HBC channel is caused due to: (i) Inter-device coupling between the transmitter and receiver, (ii) Body shadowing effects.\newline
\textbf{Inter-device coupling} occurs when the ground plate of the transmitter and the receiver are placed relatively close to each other such that they are coupled capacitively. This provides an alternate return path to the main return path formed by the floating ground plate of the device to the earth's ground $(C_{G\infty})$. This increases the effective return path capacitance thus causing channel loss to decrease.\newline
\textbf{Body shadowing effects} that rise due to fringe fields between human body and floating ground plate results from the human body acting as a conductor which is in close proximity to the floating ground plate with the air in between acting as a dielectric. The return path capacitance which completes the Capacitive HBC circuit by closing the return path between floating ground plate and the Earth's ground decreases due to being shadowed by the human body hence restricting the direct path from the floating ground plate to the earth's ground. As the fringe field and consequently the fringe capacitance from body to floating ground plate increases, the return path capacitance decreases. This has an adverse affect on the channel quality as whenever the ground plate is surrounded by the human body, the channel loss increases.

\par
Leveraging on the newly developed biophysical model in this paper, we develop a simple circuit theoretical model for the HBC channel which is used to get an expression for the channel loss in capacitive HBC. From this expression, for the first time in literature, a channel loss equation is developed as a function of the geometric parameters of the device being used as well as the position of the device with respect to the human body. This channel loss equation can be used to accurately design devices in terms of their size and position in a wearable WBAN system.\par  

The subsequent sections are organized as follows: Section II gives an overview of the working of capacitive HBC and related work done on HBC. In Section III, we look into the fringe fields and develop a biophysical model for the same. We also investigate how return path capacitance and subsequently channel loss is affected due to the presence of fringe fields. We further update the biophysical model in Section IV by studying the effects of inter-device coupling and we develop a closed form equation for the path loss in an HBC channel. We summarize and conclude our study in section V.

\section{Background and related work}

 In this study, we focus on analyzing the variability in a capacitive electro-quasistatic (EQS) HBC channel. A capacitive HBC system is where the transmitter and the receiver are coupled to the body using a single signal electrode. A floating ground electrode is used to form the closed loop using parasitic capacitances to the earth ground. The electrode on the transmitter side transmits the signal via the forward path made up by the body and is detected at the receiver end. The transmitted signals pass through the subcutaneous conductive layers in the human body. \par

 The frequency of operation for the purposes of the simulations and experiments conducted lie within the electro-quasistatic (EQS) regime \cite{NSR} of less than $1MHz$. This would ensure that the signals transmitted are constrained to stay predominantly within the body. In this frequency range, the propagation wavelength of the signal is $\geq 300 m$ which is orders of length more than the dimensions of the transmitters and receivers used $(\approx 0.03 m)$ as well as the human body $(\leq 2 m)$. \par

 Various studies have been done to characterize the human body as a channel for capacitive HBC communication. A biophysical model was proposed by Maity et al. \cite{TBME19} where measurements were used to find out ballpark values for the R-C network developed to model the human body as a capacitive HBC channel. It was shown that for capacitive HBC, the path loss in a broadband body channel is a very weak function of the distance between the receiver and transmitter. This is because the resistance being offered by the body ($R_{B}$) being negligible in comparison to the impedance due to the return path capacitance ($C_{G\infty}$) in the circuit. The path loss was also shown to be a strong function of the return path capacitance. The return path capacitance is a parasitic capacitance which forms a direct path between the ground plate of the transmitter or the receiver and the earth's ground. However, the model provided does not explore various capacitances between the device, body and the earth's ground over all positions of the devices on the human body. This doesn't explain the variability seen in an HBC channel (Fig. \ref{fig:CG_CB_Model}) due to the positional effect of the transmitter and receiver with respect to the human body and due to the close proximity between the devices. \par

 Return path capacitance was analyzed by Nath et al. \cite{TCAS2} where it was proved that the return path capacitance is a function of the device size and is independent of the height of the device from the ground plane for the radius of the device being orders of magnitude smaller than the height of the device from the earth's ground. It was also suggested that the return path capacitance value for a disc shaped device is close to the self capacitance of a disc when the height of the  disc is much larger than the dimensions of the disc. This doesn't take into consideration the effect of shadowing by the human body to reduce the return path capacitance and hence make the channel loss higher. \par
 
 Biophysical models proposed previously have characterized the human body as a constant path loss channel for capacitive HBC in the EQS regime. This study attempts to explain the variability in an HBC channel caused due to inter-device coupling and the body to floating ground plane coupling. The reduction in return path capacitance due to the effects of body shadowing is also explained. Simulation results which are further backed up by measurements carried out in low frequencies $(\leq 1 MHz)$ are used to validate the proposed advanced biophysical model.

\begin{figure}[h!]
\centering
\includegraphics[width=0.5\textwidth]{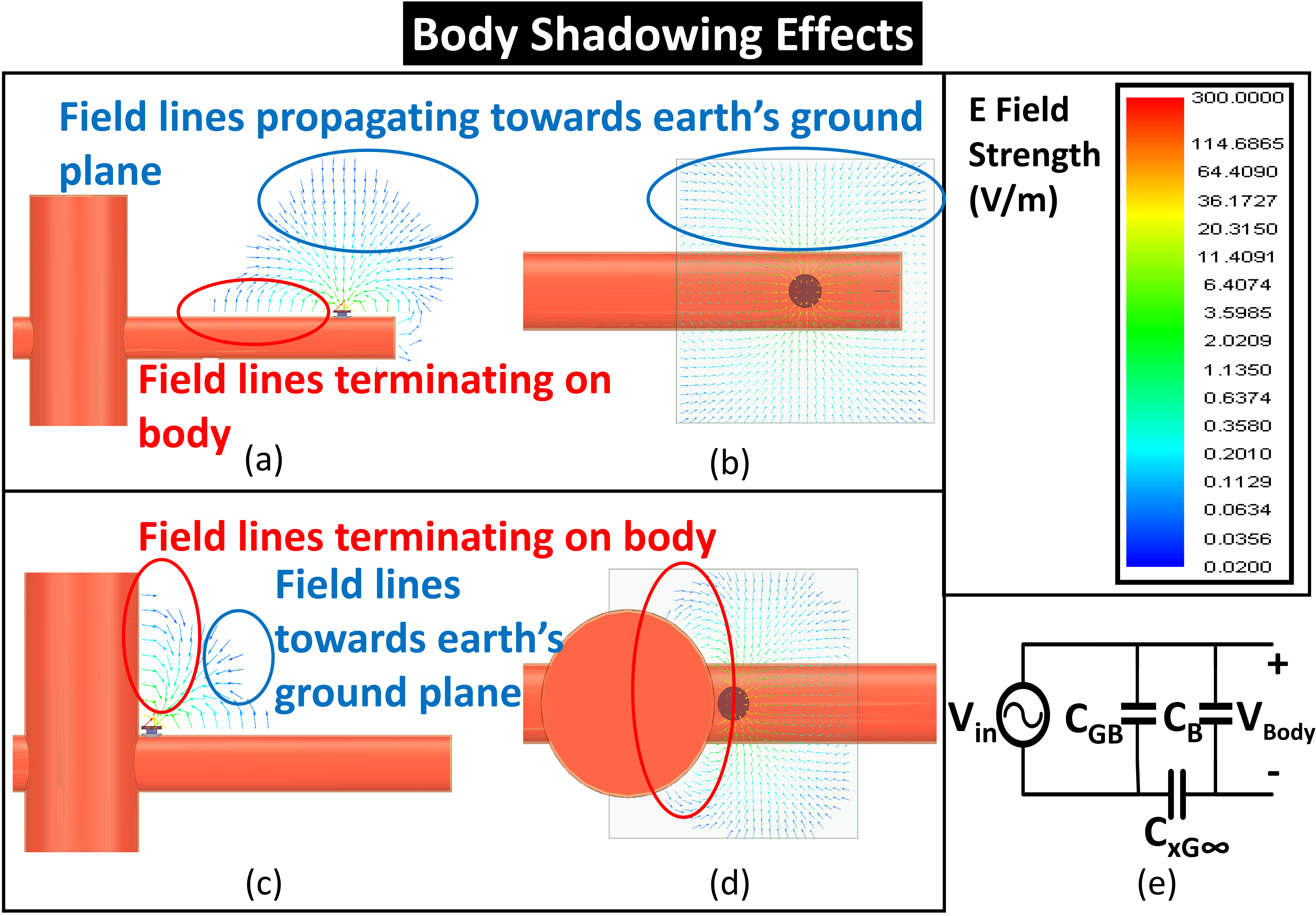}
\caption{Field line study using Ansys HFSS showing electric field line patterns. The transmitter when placed away from the torso has lesser number of field lines terminating on the body as observed from (a) the front view and (b) the top view. When the transmitter is brought closer to the torso, most of the field lines terminate on the torso. This is illustrated in (c) from the front view and (d) from the top view. (e) Simplified circuit model developed for a capacitive HBC channel with respect to the amount of signal being coupled to the body.}
\label{fig:fringe_field_plot}
\end{figure}

\section{Biophysical model for body shadowing effects}

The channel loss for a capacitive HBC system is dependent on the return path capacitance. A thorough understanding of the return path capacitance becomes important in the study of the channel characteristics for capacitive HBC. It is therefore essential to find an accurate theoretical model for the return path capacitance which will further help us in developing an efficient model for the capacitive HBC channel.

\subsection{Self capacitance of a disc}

The transmitter and receiver ground plates for a wearable device can be modelled as a disc. The radius of the disc for wearable devices is typically in the range of a few centimeters. This means that the device ground is more than $10x$ smaller than the height of the device from the earth's ground. For these typical ground sizes, the capacitance of the ground plane is approximately equal to the self capacitance of the device which is dependent upon the geometry of the disc \cite{Disc_Self_Cap} \cite{TCAS2}. This allows us to express the return path capacitance as a function of the radius of the disc $a$ and its height $h$ as shown below:
\begin{equation}
    C_{disc} = 8\epsilon_0a\left[1+0.87\left(\frac{h}{2a}\right)^{0.76}\right]
\end{equation} 
For a thin disc with $a<<h$:
\begin{equation}
    C_{disc} = 8\epsilon_0a
\end{equation}
\par
Fig. \ref{fig:CG_CB_Circuit} shows the simplified circuit model for the measurement of body potential when a voltage source is connected to the human body. The biophysical model shown in Fig. \ref{fig:CG_CB_Circuit} (a) is a simplification based on the premise that the tissue resistance of the body (resistance offered by the body inside the skin layer) is of the order of a few $100\Omega s$ \cite{Rtissue}. Further, the skin layer has an impedance (skin resistance) of the order of $10k \Omega s$ \cite{Rskin}. In contrast, the impedance provided by the return path capacitance $(C_{G\infty} \approx 1pF)$ is of the order of greater than $100k \Omega$ for the operating frequency of less than $1MHz$ (EQS region) which is an order of magnitude more. Due to low impedance provided by the tissue layer of the body, the whole body appears to be shorted and therefore acts like a conductor. This allows us to neglect the skin and tissue resistance which has been clubbed together into the component shown as $R_{Body}$ in Fig. \ref{fig:CG_CB_Circuit}(a) and (b). The voltage transfer function for the given circuit model (Fig. \ref{fig:CG_CB_Circuit} (c)) can be written as follows:
\begin{equation}
    \frac{V_{Body}}{V_{in}} \approx \frac{C_{G\infty}}{C_{B}}
        \label{Eqn:Body_Potential}
\end{equation}
This can be expressed in terms of the geometry of the device as:
\begin{equation*}
    \frac{V_{Body}}{V_{in}} \approx \frac{8\epsilon_0a}{C_{B}}
\end{equation*}

\subsection{Effect of body on the return path capacitance - Fringe fields}

Field lines from the body don't terminate only to the earth's ground but also to the transmitter ground plate as shown by the Ansys HFSS EM simulation in Fig. \ref{fig:fringe_field_plot}. Hence, the simple model shown in Fig. \ref{fig:CG_CB_Circuit} does not completely characterize the human body channel as it doesn't take into consideration the coupling capacitance between the body and the transmitter ground plate. Fig.  \ref{fig:fringe_field_plot} (a) and (b) illustrates how the field lines are distributed when the transmitter is placed away from the torso, close to the wrists. We see that most of the field lines in this case don't terminate on the body. This is because the transmitter is away from the torso which has a higher surface area than the arms and the body doesn't effectively block the path from the transmitter's ground plate to the earth's ground. However, when the transmitter is shifted close to the torso as shown in Fig. \ref{fig:fringe_field_plot} (c) and (d), majority of the field lines terminate on the body rather than terminating to the earth's ground. The torso now being close to the floating ground plate of the transmitter shadows the device thereby blocking the field lines from reaching the earth's ground. \par

The body to floating ground plate capacitance $(C_{GB})$ results from the fringe fields between the transmitter ground and the body ($C_F$) as well as the parallel plate capacitance between the signal plate and the ground plate ($C_{PP}$). The presence of fringe capacitance reduces the return path capacitance hence increasing the channel loss. This phenomenon has been discussed in detail in the Supplemental Material where a sphere thought model (Fig. S1) is used to describe how the fringe capacitance affects the value of the return path capacitance.

\begin{figure}[h!]
\centering
\includegraphics[width=0.5\textwidth]{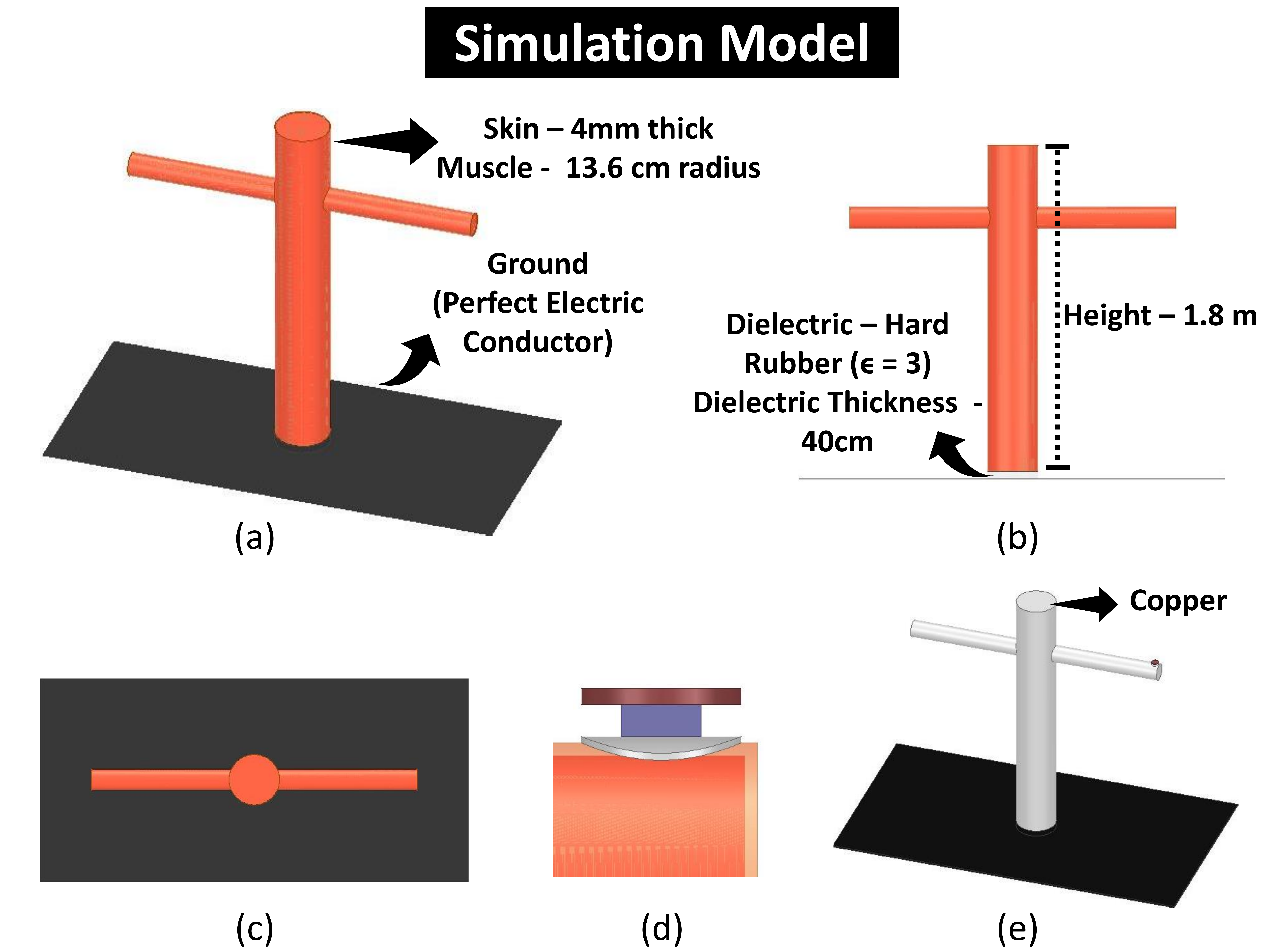}
\caption{(a) Simplified human model made with two cross connected cylinders consisting of a 4mm thick skin shell and muscle. Front view (b) and top view (c) of the model is illustrated. (d) The structure of a transmitter and receiver device used. The top disc is the floating ground plate where as the bottom disc is the signal plate. (e) Model used for Maxwell simulations. The cross connected cylindrical model is made up of copper. }
\label{fig:HFSS}
\end{figure}

\subsection{Advanced biophysical model}

The body to floating ground plate capacitance is modeled as $(C_{GB})$. In the model shown, $C_{GB}$ can be further broken down into two separate components: plate-to-plate capacitance $(C_{PP})$ and fringe capacitance $(C_F)$.

\begin{equation*}
    C_{GB} = C_{PP} + C_F
\end{equation*}

Plate-to-plate capacitance is the parallel plate capacitance between the signal plate and the ground plate of the transmitter which can be mathematically represented as:

\begin{equation}
    C_{PP} = \frac{\epsilon_0 \pi a^2}{t}
    \label{eqn:Cpp}
\end{equation}
where, $a$ is the radius of the circular disc which models the floating ground plate and $t$ is the thickness of the device.\newline
Fringe capacitance is present between the human body and the ground plate of the device. The resulting circuit model for the HBC channel is shown in Fig. \ref{fig:fringe_field_plot} (e). The addition of floating ground plate to body capacitance doesn't directly affect the value of body potential $(V_{Body})$ observed from the circuit. However, due to the effect of body shadowing, some of the field lines which were going to the earth's ground instead terminate on the body (further discussed in Supplementary Material, Fig. S1). This reduces the return path capacitance between the floating ground plate of the device and the earth's ground ($C_{xG\infty}$). This can be written mathematically as: 
\begin{equation*}
    C_{xG\infty} \leq 8\epsilon_0 a
\end{equation*}
where, a is the radius of the disc used as the transmitter ground. 
The return path capacitance can be further represented as: 
\begin{equation}
    C_{xG\infty} = x \times C_{G\infty}
    \label{Eqn:Cx_less_Cg}
\end{equation}
where $x$ is a real valued number lying between 0 and 1 depending upon the position of the device with respect to the body.

\begin{figure*}[t!]
\centering
\includegraphics[width=\textwidth]{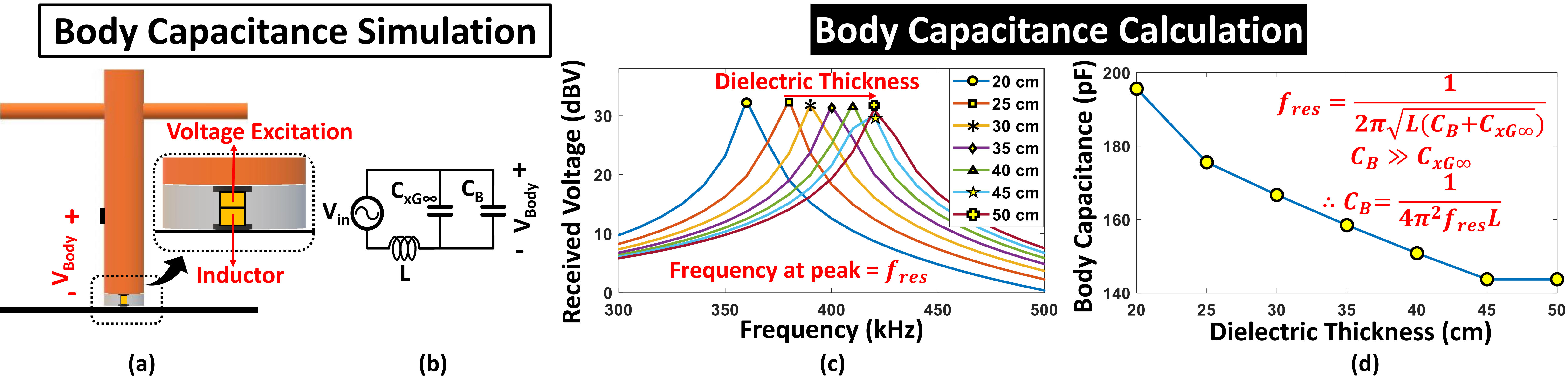}
\caption{(a) Simulation setup for body capacitance simulation with the transmitter placed between the body and the earth's ground. A voltage excitation is applied with an inductor of $1mH$ in series at the transmitter side. (b) Circuit model for the simulation setup showing a simple parallel LC-resonant circuit. (c) Resonant peaks for various dielectric thickness values. (d) Body Capacitance values for changing dielectric thickness shows how decreasing dielectric thickness increases the body to ground capacitance value which in turn increases the channel loss.}
\label{fig:body_Cap_Graph}
\end{figure*}

\subsection{Simulation setup}

The simulations are carried out using Ansys High-Frequency Structure Simulator (HFSS) and Maxwell which are Finite Elements Analysis (FEA) based EM solvers. HFSS is an EM solver for high-frequency applications whereas Maxwell is used in our simulations as an electrostatic solver.

For the simulations, we use a simplified crossed cylindrical model of the human body as described by Fig. \ref{fig:HFSS}. Even though the model may seem to be an oversimplification of the human body, previous studies by Maity et al. \cite{Safety_Study} observed the electric and magnetic field field distribution in and around the human body for HBC using EM simulations. These results compared the field distributions between the simplified cross cylindrical model with a complex human body model - VHP Female v2.2 \cite{neva_model} from Neva Electromagnetics. The field distributions are observed to be identical inside and outside the model. These results make sense because on checking the dielectric permittivity of various tissues on the body \cite{Gabriel_1996}, it can be seen that most of the human tissues (with the exception of blood) have a higher permittivity than air. Further, it can be seen that most of the tissues have a low but non-negligible conductivity. When compared to the electrical properties of air (relative permittivity of 1 and conductivity of 0), the body acts like a homogeneous mass of high dielectric permittivity and low but positive conductivity. As the HBC devices we will be looking into works at the interface of air and body, the simplified human body model provides us with realistic results without any loss of generality. Hence, the simplified cross cylindrical model provides us with dependable results and further reduces computational complexity and time by orders of magnitude for simulations performed over multiple frequency ranges.\par 
The ground plane in the model acts as a plane of reference for the simulations to calculate the potential on the body. A dielectric "Hard Rubber" of dielectric constant of $3$ is present between the body and the ground plane. The dielectric helps in modeling the body capacitance $(C_B)$ present for the simplified cross cylindrical model. Changing the height of the dielectric changes the capacitance between the body and the earth's ground plane.\par
The dielectric and conductive properties of muscle and skin in the model has been adapted from Gabriel et al. \cite{Gabriel_1996}. The structure of the transmitter and receivers used for capacitive HBC is shown in Fig. \ref{fig:HFSS} (d), where one of the electrodes is the signal plane while the other acts as a floating ground plane.  

In a Maxwell solver, any object with a conductivity of less than 1 is considered to be a dielectric. However, in an EQS paradigm, it has been shown that human body acts like a conductor \cite{JSSC}\cite{CICC}. As the dielectric properties of human muscle and skin are a function of the frequency of use, Maxwell solver may consider the body to be a dielectric instead of a conductor in some cases. To ensure that the solver considers the human body to be a conductor, we change the material of the human body to a conductor (for example, copper) as shown in Fig. \ref{fig:HFSS} (e). We use Maxwell solver as a tool for calculating capacitance between two entities. The capacitance between two conductors is a function of the structure and shape of the conductors, the distance between the two conductors and the relative permittivity of the dielectric between them. As the capacitance is not a function of the conductivity of the conductors used, the use of copper as a material instead of human muscles and skin doesn't compromise the results of the simulations.    

\subsection{Simulation results}

\subsubsection{Body Capacitance $(C_{B})$}

Body capacitance is the lumped capacitance present between the human body and the earth's ground which is an important component in the circuit modelling of the HBC channel as shown in Fig. \ref{fig:fringe_field_plot} (e). Body capacitance is calculated with reference to a ground plane using the simulation setup shown in Fig. \ref{fig:body_Cap_Graph} (a). The transmitter device is placed between the earth's ground plane and the body. A voltage excitation is applied in series with a lumped inductance of $1mH$.  Body capacitance is obtained for various heights of the dielectric between the body and the earth's ground. The circuit model of the simulation setup is presented in Fig. \ref{fig:body_Cap_Graph} (b) which is a simple parallel L-C resonant circuit. The value of the inductor must be chosen carefully as it should be ensured that the resonant peaks obtained lie in the EQS region of less than $1 MHz$. Reducing the inductance to $1\mu H$ will move the resonant peaks to a frequency of more than $10 MHz$ where the peaks may be falsely represented due to the body acting as a more efficient antenna at that frequency range.\par
The resonant peaks obtained for varying dielectric heights obtained are shown in Fig. \ref{fig:body_Cap_Graph} (c). We see that as the dielectric thickness increases, the body capacitance value reduces. Fig. \ref{fig:body_Cap_Graph} (d) gives us the body capacitance values for changing dielectric thickness. For our simulations a dielectric thickness of $40cm$ is chosen with a body capacitance of $150.838 pF$. 

\begin{figure}[h!]
\centering
\includegraphics[width=0.4\textwidth]{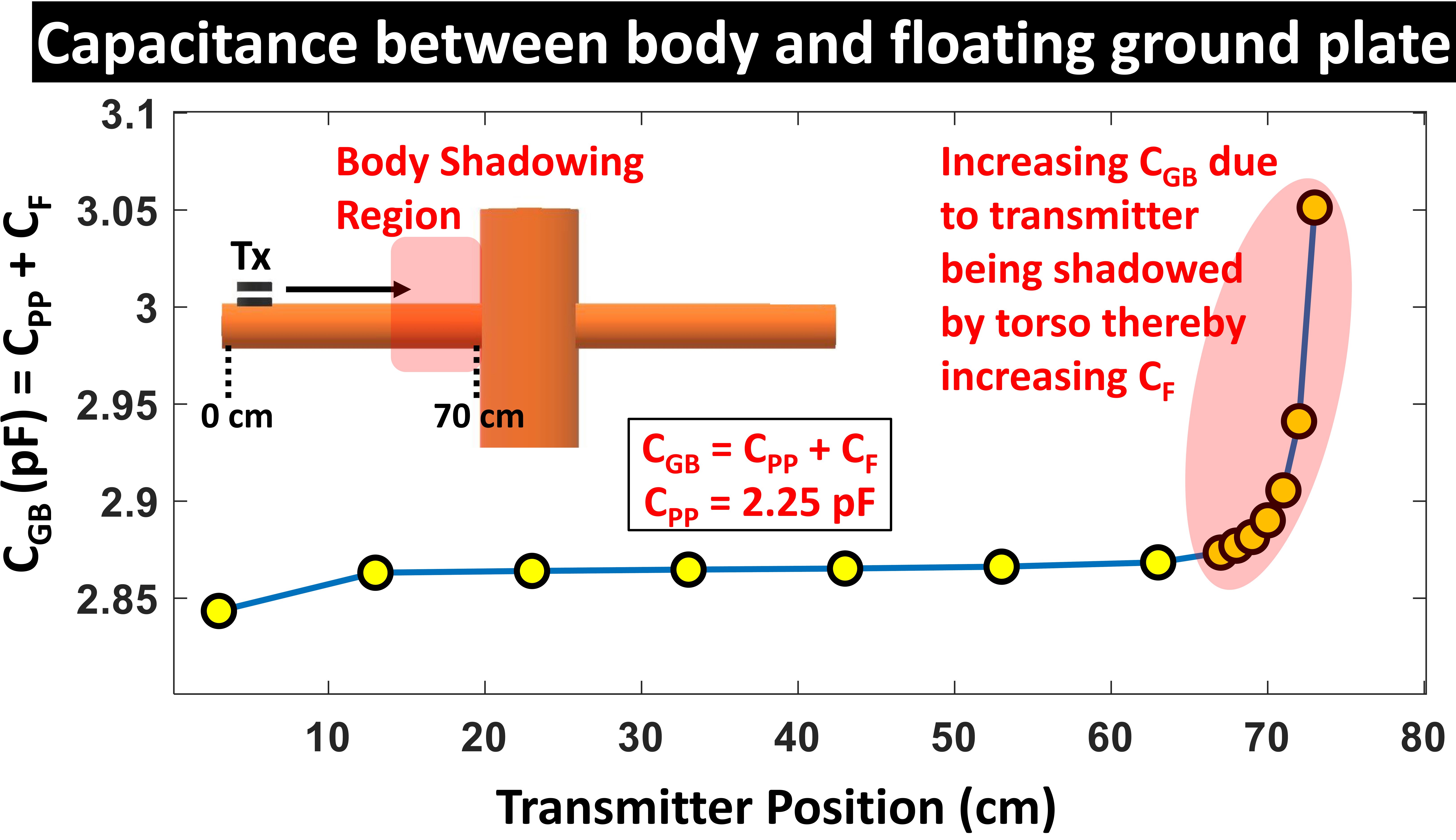}
\caption{Body to floating ground plate capacitance ($C_{GB}$) versus position of the transmitter on the arm. On placing the transmitter closer to the torso, the body shadows the floating ground plate more subsequently increasing the fringe capacitance $(C_F)$ thereby increasing $C_{GB}$.}
\label{fig:fringe_Cap_Graph}
\end{figure}
\subsubsection{Capacitance between floating ground plate and body $(C_{GB})$}
Fringe fields present between body and floating ground plate of the device (Fig. \ref{fig:fringe_field_plot}) effectively contributes in reduction of the return path capacitance due to the effect of stealing field lines that were terminating to the earth's ground. These fringe fields are modeled in the biophysical circuit model as the body to floating ground plate capacitance $(C_{GB})$. The capacitance between human body and the transmitter's floating ground plate is shown in Fig. \ref{fig:fringe_Cap_Graph}. The transmitter is moved across the arm while Maxwell electrostatic solver is used to calculate the capacitance between the floating ground plate and the human body. We observe that as the transmitter moves closer to the torso, the fringe capacitance $(C_F)$ increases thus increasing the body to floating ground plate capacitance $(C_{GB})$. The plate-to-plate capacitance $(C_{PP})$ is found out using Maxwell solver. For the given simulation $C_{PP}$ is constant at $2.25 pF$ as the area of the plates and the thickness of the dielectric for the transmitter is unchanged.

\subsubsection{Received potential on the body}
\begin{figure}[h!]
\centering
\includegraphics[width=0.4\textwidth]{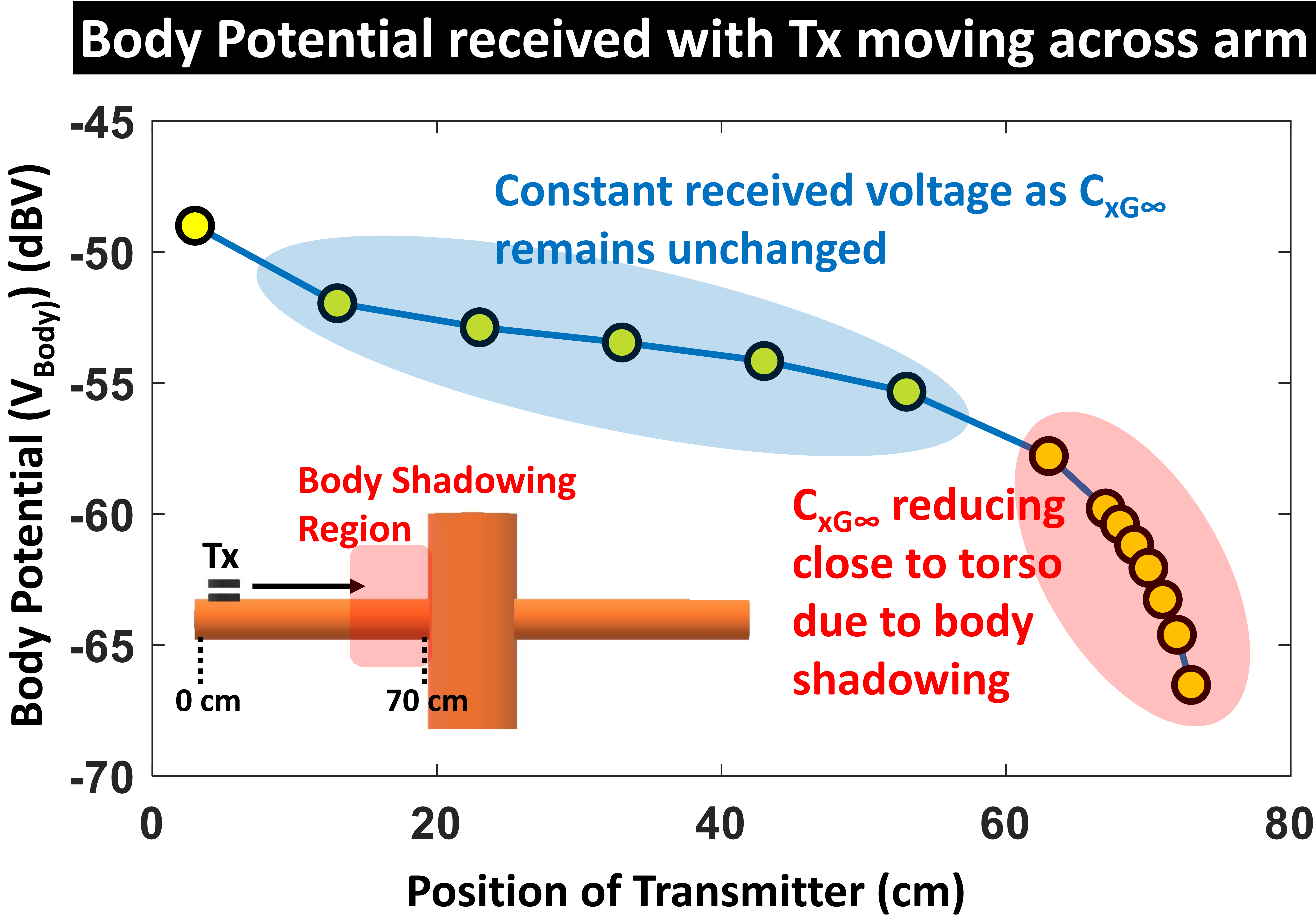}
\caption{Potential received on the body Varying with position of the transmitter on the arm. Body potential drops when the transmitter moves closer to the torso}
\label{fig:Body_Potential_Graph}
\end{figure}

The received voltage on the human body is seen to be a strong function of the return path capacitance. Fig. \ref{fig:Body_Potential_Graph} illustrates how the received potential on the body changes when the transmitter moves across the arm. The received voltage is constant when the transmitter is positioned away from the torso. However, the body potential decreases rapidly as the transmitter moves close to the torso. 
\begin{figure*}[t!]
\centering
\includegraphics[width=\textwidth]{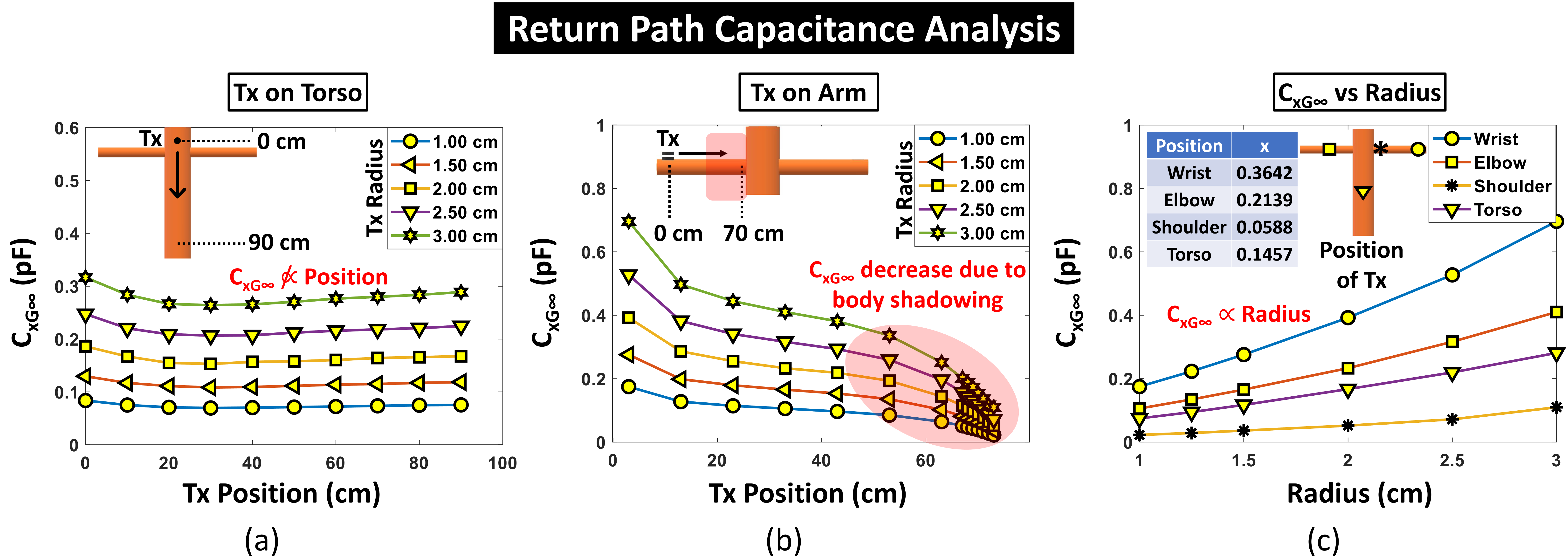}
\caption{Variation of the return path capacitance is shown when the transmitter (Tx) position is varied along (a) torso, (b) arm, for changing device radius. (c) Return path capacitance versus radius is plotted. Return path capacitance is shown to be directly proportional to the radius of the device across all positions on the body. The table shows typical values of $x$ for the positions shown in the graph.}
\label{fig:Return_Path_Cap_Graph}
\end{figure*}
\subsubsection{Return Path Capacitance $(C_{xG\infty})$}
Return path capacitance originally assumed to be the self capacitance of the disc \cite{TCAS2} has been shown to reduce due to shadowing by the human body. This reduction is analyzed using results found in previous sections and new HFSS simulations so as to find an equation that accurately models the capacitance. \newline
The return path capacitance for the transmitter is calculated from the result obtained for received potential on the body (Fig. \ref{fig:Body_Potential_Graph} and the body capacitance values (Fig. \ref{fig:body_Cap_Graph} (d)). Eqn: \ref{Eqn:Body_Potential} is rectified to include the parameter $C_{xG\infty}$ instead of $C_{G\infty}$ $(=8\epsilon_0a)$ to accommodate the reduction in return path capacitance due to disc being shadowed by the human body. Return path capacitance is calculated as shown:

\begin{equation*}
    \frac{V_{Body}}{V_{in}} \approx \frac{C_{xG\infty}}{C_{B}}
\end{equation*}

\begin{equation}
    C_{xG\infty} = C_{B} \times  \frac{V_{Body}}{V_{in}}
\end{equation}
For a $40cm$ dielectric:
\begin{equation*}
    C_{xG\infty} = 150.838 (pF) \times  \frac{V_{Body}}{V_{in}}
\end{equation*}

Fig. \ref{fig:Return_Path_Cap_Graph} (a) shows that the return path capacitance of the transmitter stays constant for a particular radius when the device is placed on the torso. We also observe that there is an increase in the return path capacitance for the corner case where the transmitter is placed close to the head where the shadowing effects of the body are comparatively less than other parts of the torso. \par 
Fig. \ref{fig:Return_Path_Cap_Graph} (b) illustrates how return path capacitance varies with the position of the transmitter when the device is placed on the arm. We observe that the return path capacitance decreases when the transmitter is close to the shoulder. This is because the body blocks the direct path between the ground plate of the transmitter and the earth's ground more as the device is closer to the torso. Further, we observe that the return path capacitance increases when the device is moved away from the torso. The body's shadowing effects are least felt when the transmitter is placed away from the torso near the wrists where the return path capacitance value is observed to be highest. \par
Fig. \ref{fig:Return_Path_Cap_Graph} (c) shows how the return path capacitance varies with radius of the floating ground plate for different positions on the human body. The figure confirms our hypothesis made in Eqn: \ref{Eqn:Cx_less_Cg} showing that the return path capacitance formed is a fraction of the self capacitance of a disc. This can be represented as:
\begin{equation}
     C_{xG\infty} = x \times 8\epsilon_0 a
     \label{Eqn:return_path_cap}
\end{equation}
 where $x < 1$.\par
 The value of $x$ will depend upon the position of the floating ground plane with respect to the human body.

\section{Biophysical modeling of Inter-device Coupling}

\subsection{Circuit modelling of an HBC receiver}

In the previous section, we have dealt with the HBC channel from the perspective of how much signal gets coupled to the human body. However, the channel loss for an HBC system will be incomplete without considering the structure and position of the receiver and how it affects the communication channel. \par
\begin{figure}[b!]
\centering
\includegraphics[width=0.5\textwidth]{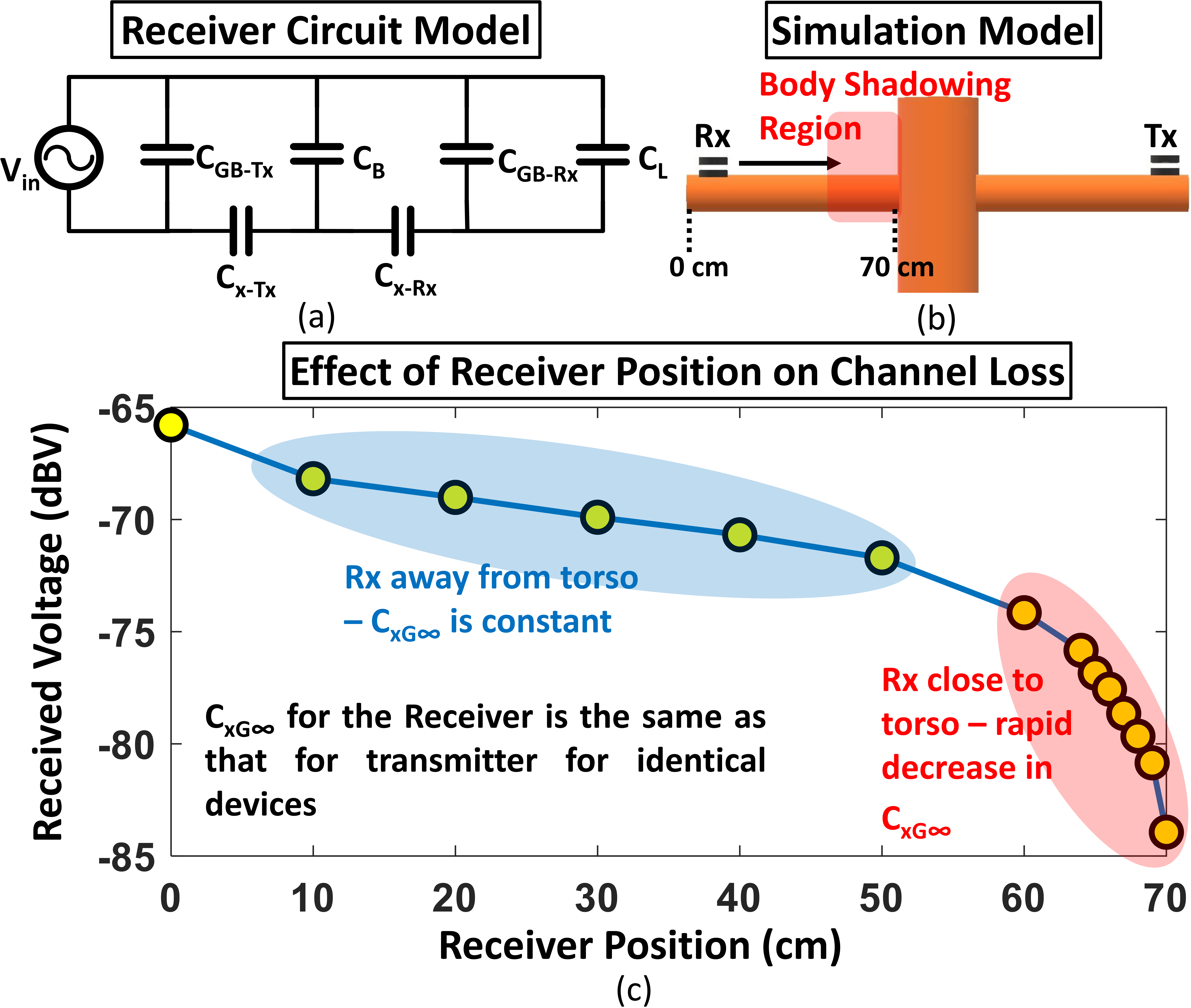}
\caption{a) Circuit modelling for a capacitive HBC system with a receiver device. Channel variability due to body shadowing effects are captured in this circuit model. (b) Receiver and transmitter position with respect to the human body. (c) Received voltage versus receiver position on the body. The received voltage shows a similar trend as shown by a moving transmitter in Fig. \ref{fig:Body_Potential_Graph}. This indicates that the return path capacitance at the receiver end also varies with the position of the receiver}
\label{fig:Rx_Circuit}
\end{figure}

Figure \ref{fig:Rx_Circuit} (a) shows how the receiver changes the circuit modeling of the HBC channel. The receiver device is similar in structure to that of the transmitter and hence we get a capacitive coupling between the body and the floating receiver ground plate $(C_{GB-Rx})$. Return path capacitance is now present from the transmitter floating ground plate to the earth's ground $(C_{x-Tx})$ as well as from the receiver floating ground plate to the earth's ground $(C_{x-Rx})$. The received voltage is read across a load capacitance $(C_L)$. \par

The return path capacitance for the receiver $(C_{x-Rx})$ is characterized in the same way as it was done for the transmitter. This is because the return path capacitance is shown to be a function of the geometry of a device and the receiver is structurally similar to the transmitter device. This would mean that the return path capacitance will also be a strong function of the radius of the disc and the position in which the receiver is placed.\par

Received voltage across the load capacitance $(V_o)$ can be represented as:

\begin{equation*}
    \frac{V_{o}}{V_{in}} = \frac{C_{x-Tx}}{C_{B}} \times \frac{C_{x-Rx}}{C_{x-Rx}+C_{GB-Rx}+C_L}
\end{equation*}
As the return path capacitance at the receiver side $(C_{x-Rx} < 1pF)$ is an order of magnitude smaller than the sum of load capacitance and the capacitance between body and floating ground plate of the receiver $(C_L + C_{GB} \sim 10pF)$, we can approximate the above equation as:
\begin{equation}
    \frac{V_{o}}{V_{in}} = \frac{C_{x-Tx}}{C_{B}} \times \frac{C_{x-Rx}}{C_{GB-Rx}+C_L}
    \label{Eqn:Rx_Potential}
\end{equation}

\begin{figure*}
\centering
\includegraphics[width=0.9\textwidth]{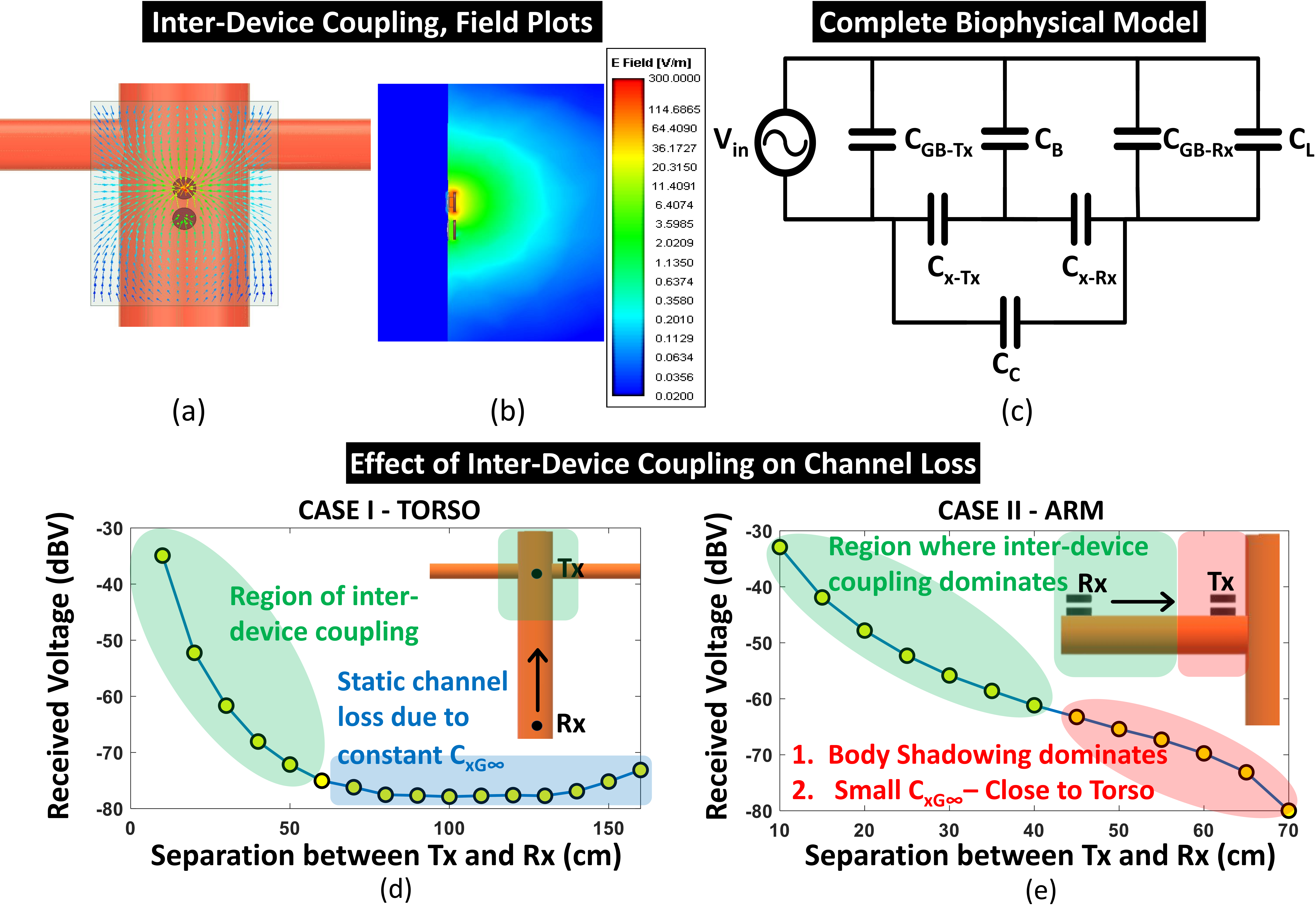}
\caption{Field line plots (a) and the electric field distribution (b) illustrates that electric field lines of high magnitude emanating from the transmitter terminate on the receiver when the devices are placed close to each other. (c) Advanced bio-physical circuit model for a capacitive HBC channel including channel variability.  Variation of received voltage with respect to separation between transmitter and receiver is studied with the devices placed on (d) torso and (e) arm.}
\label{fig:Inter_Device_Coupling_Combined}
\end{figure*}

\subsection{HFSS results to verify circuit model}

Fig. \ref{fig:Rx_Circuit} (c) shows how the position of a receiver affects the channel loss. For this simulation, the transmitter is kept stationary on one arm ensuring that the return path capacitance on the transmitter side $(C_{x-Tx})$ is constant. The receiver is moved along the opposite arm so that the receiver and transmitter are far from each other to avoid any other second order effects. \par

The trend seen in Fig. \ref{fig:Rx_Circuit} (c) resembles the body potential observed when the transmitter is moved across the arm in Fig. \ref{fig:Body_Potential_Graph}. This is because the receiver changes position in turn changing the return path capacitance of the receiver $(C_{x-Rx})$. The variation of $C_{x-Rx}$ will be the same as that obtained in Fig. \ref{fig:Return_Path_Cap_Graph} for a specific radius as the receiver ground plane is structurally similar to that of the transmitter.

\subsection{Inter-device coupling in EQS HBC}

For all the simulations performed so far, we maintained a large separation $(>50cm)$ between the transmitter and the receiver. This was to ensure that the return path for the capacitive HBC system is formed only through the earth's ground. However, when the separation between transmitter and receiver is reduced, the floating ground plane of the transmitter and receiver couple to form an alternate return path for the capacitive HBC system. This increases the effective return path capacitance thereby decreasing the channel loss of the system. \par 
Fig. \ref{fig:Inter_Device_Coupling_Combined} (a) and (b) shows the electric field lines and the distribution of electric field close to the transmitter. The field lines with higher magnitudes terminate on the floating ground plate of the receiver when the devices are close to each other. This creates a capacitive path between the ground plate of devices that are in close proximity. As the receiver moves away from the transmitter, the magnitude of electric field around the receiver reduces thus increasing the channel loss. \par

As illustrated in Fig. \ref{fig:CG_CB_Model}, a capacitive path forms between the floating ground plates of devices that are close to each other. The return path capacitance now is not just via the earth's ground plane but also between the two floating ground plates of the devices thus increasing the return path capacitance of the system. This phenomenon is termed as inter-device coupling. Inter-device coupling reduces the path loss of the capacitive HBC channel significantly when the transmitter and receiver are close to each other.

\subsection{Updated biophysical model of HBC}

Inter-device coupling between the transmitter and receiver floating ground plates is modeled in our biophysical circuit model with a coupling capacitance $(C_c)$. Fig. \ref{fig:Inter_Device_Coupling_Combined} (c) shows the complete biophysical model capturing the channel variability for a capacitive HBC system working in EQS domain. \par

Received voltage across the load capacitance can be mathematically represented as shown in Eqn: \ref{eqn:Inter_Device_Coupling_Full}. Since the self capacitance of the body $(C_B \approx 150pF)$ and the load capacitance $(C_L \approx 10pF)$ is orders of magnitude larger than the return path capacitance $(C_{xG\infty} < 1pF)$, the equation can be simplified to:
\begin{equation}
    \frac{V_{o}}{V_{in}} = \frac{C_c+ \left(\frac{C_{x-Rx}C_{x-Tx}}{C_B}\right)}{C_c + C_L + C_{GB-Rx}}
    \label{eqn:Inter_Device_Coupling}
\end{equation}

\begin{strip}
\noindent
\fbox{
    \parbox{\textwidth}{%
    \begin{equation}
         \frac{V_{o}}{V_{in}} = \frac{C_c[C_{B}+C_{x-Rx}+C_{x-Tx}]+C_{x-Rx}C_{x-Tx}}{C_c[C_{B}+C_{x-Rx}+C_{x-Tx}]+(C_{B}+C_{x-Rx})(C_L + C_{GB-Rx}+C_{x-Tx})+C_{x-Tx}(C_L + C_{GB-Rx})}
         \label{eqn:Inter_Device_Coupling_Full}
    \end{equation}
    }}%
\end{strip}

The coupling capacitance $(C_c)$ is typically less than $1 fF$ when the two devices are kept away from each other $(>50cm)$ which is orders of magnitude less than the return path capacitance. In such a situation, the equation can be then simplified to: 

\begin{equation}
    \frac{V_{o}}{V_{in}} = \frac{C_{x-Rx}C_{x-Tx}}{C_B (C_L+C_{GB-Rx}) }
    \label{eqn:inter_device_coupling_simplified}
\end{equation}

which is the same as the path loss equation obtained for the circuit model in Fig. \ref{fig:Rx_Circuit} (a) (Eqn: \ref{Eqn:Rx_Potential}). The variation of body to floating ground plate capacitance $C_{GB}$ is shown in Fig. \ref{fig:fringe_Cap_Graph}. The variation in $C_{GB}$ when the device moves closer to the torso is caused due to an increase in the fringe capacitance $C_F$.

This increase in fringe capacitance varies the body to floating ground plate capacitance from $2.85 pF$ to about $3 pF$. This variation in $C_{GB}$ is negligible as the percent change in body to floating ground plate capacitance due to increase in fringe capacitance is very low$(\leq 5\%)$.

Further, the variation of fringe capacitance is negligible compared to the load capacitance $(C_L)$ which is typically in the order of $10 pF$ for small devices. Hence, the body to floating ground plate capacitance $(C_{GB-Rx})$ can be considered to be constant over different positions on the human body. \par

When two devices come close to each other, the coupling capacitance value increases. The value of coupling capacitance in Fig. \ref{fig:Inter_Device_Coupling_Combined} (d) and (e) for a separation of 10cm between two devices is around $60 fF$. Further, the coupling capacitance is of the order of $10s$ of femtofarads for small separations $(\leq 20 cm)$ between the two devices. Hence, from Eqn: \ref{eqn:Inter_Device_Coupling} it can be inferred that the contribution of return path capacitances $(C_{x-Tx}, C_{x-Rx})$ which is of the order of 100s of femtofarads can not be neglected when the devices are placed close to each other.

\subsection{Simulation results}

The effects of inter-device coupling on the channel characteristics have been studied separately for the two cases: 
\begin{itemize}
    \item Transmitter and receiver placed on the torso
    \item Transmitter and receiver placed on the same arm
\end{itemize}

\par
For the first scenario, the transmitter device is kept at the same position on the torso while the receiver is brought closer to the transmitter as shown in Fig. \ref{fig:Inter_Device_Coupling_Combined} (d).

Since the two devices are placed on the torso, body shadowing is present under all situations. The reduction in return path capacitance due to shadowing is constant all over the torso as shown in Fig. \ref{fig:Return_Path_Cap_Graph} (a). Inter-device coupling starts affecting the channel loss only when the two devices are in close proximity to each other which is marked by the region shaded green in Fig. \ref{fig:Inter_Device_Coupling_Combined}(d).
The channel loss is shown in Fig. \ref{fig:Inter_Device_Coupling_Combined} (d) for varying separation between the devices. For a separation of more than $50cm$, we observe that the channel loss is constant as shown by the region shaded blue in Fig. \ref{fig:Inter_Device_Coupling_Combined} (d). However, when the separation between transmitter and the receiver reduces to less than $50cm$, we observe that the channel loss starts reducing. The reduction in channel loss is seen to be inversely proportional to the separation between transmitter and receiver. \par

For the second case, the transmitter and receiver are both on the arm where the transmitter is kept stationary while the receiver is placed on the same arm and is moved towards the transmitter as shown in Fig. \ref{fig:Inter_Device_Coupling_Combined} (e). We observe that two different effects, change in return path capacitance with position of the receiver due to shadowing by torso and inter-device coupling between transmitter and receiver add up to give us the resulting plots. When the receiver is close to the torso, body shadowing dominates inter-device coupling. Here, the channel loss increases due to rapid reduction in return path capacitance close to the torso. This is illustrated by the region shaded by red in Fig. \ref{fig:Inter_Device_Coupling_Combined}(e). Here, for a separation of about $45 cm$ to $70 cm$ between the two devices, the receiver is close to the torso and the effect of inter-device coupling is overpowered by body shadowing effects. \newline
As the receiver is moved away from the torso and close to the transmitter, inter-device coupling capacitance starts increasing. When the separation becomes lesser than $40 cm$, inter-device coupling dominates body shadow effects as the torso is ineffective in blocking the path between device and earth's ground when the receiver is close to the elbow of the model.  This is shown by the region shaded as green where for a separation of less than $40 cm$, we observe rapid increase in received voltage. Hence, we observe that channel loss starts decreasing due to increase in inter-device coupling capacitance which provides another return path for the system thus increasing the effective return path capacitance.\par
The flat band region shown in the graph in Fig. \ref{fig:CG_CB_Model} is not present in the simulation results shown in Fig. \ref{fig:Inter_Device_Coupling_Combined} (e) as the length of the arm is small which causes overlap between the regions where inter-device coupling and body shadow effects cause variability in the channel loss. 

\begin{figure}[h!]
\centering
\includegraphics[width=0.5\textwidth]{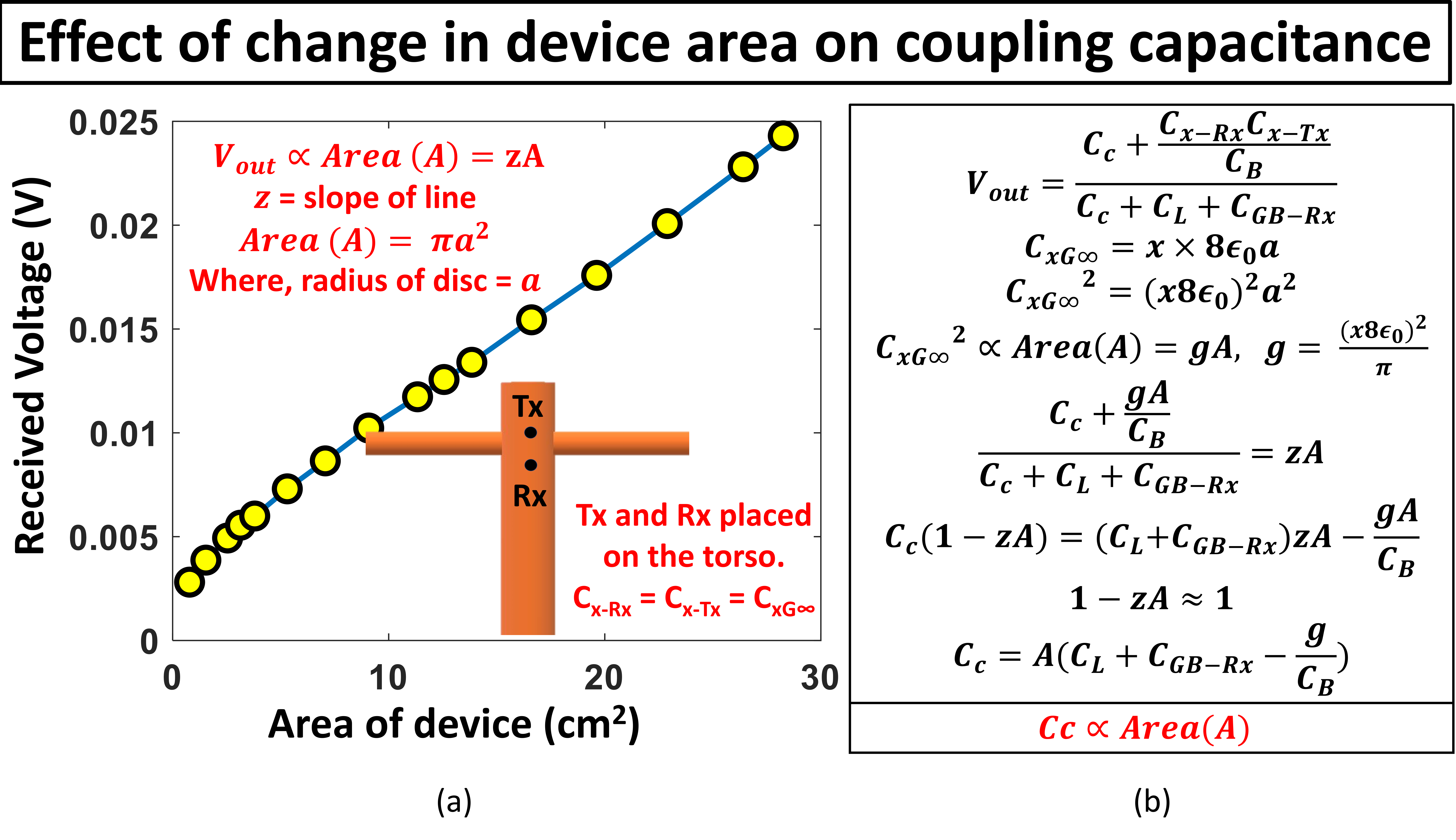}
\caption{(a) Received voltage is observed as a function of the area of device. The transmitter and receiver devices are  identical. The received voltage is shown to be proportional to the area of the device. (b) Inter-device coupling capacitance is shown to be proportional to the area of the device}
\label{fig:Area_Inter_Device_Coupling_graph}
\end{figure}

To explore the effect of change in device area on the inter-device coupling capacitance, we place the transmitter and the receiver close to each other as illustrated in Fig. \ref{fig:Area_Inter_Device_Coupling_graph} (a). Position of the devices are kept unchanged while the radius of the devices are varied to observe the received voltage. The graph shown in Fig. \ref{fig:Area_Inter_Device_Coupling_graph} (a) represents how the received voltage varies with the area of the device. We see that the received voltage increases linearly with increase in area of the transmitter and receiver devices. Fig. \ref{fig:Area_Inter_Device_Coupling_graph} (b) shows that the result obtained in the previous graph implies that the inter-device coupling capacitance linearly increases with area of the plates. Therefore, we can model the coupling capacitance $(C_c)$ mathematically as:

\begin{equation}
    C_c = \frac{kA}{d}= k\frac{\pi a^2}{d}
    \label{Eqn:Coupling_Cap}
\end{equation} 
where $A$ is the area of the device, $d$ is the distance between the transmitter and receiver device and $k$ is the constant of proportionality.

\begin{figure}[b!]
\centering
\includegraphics[width=0.5\textwidth]{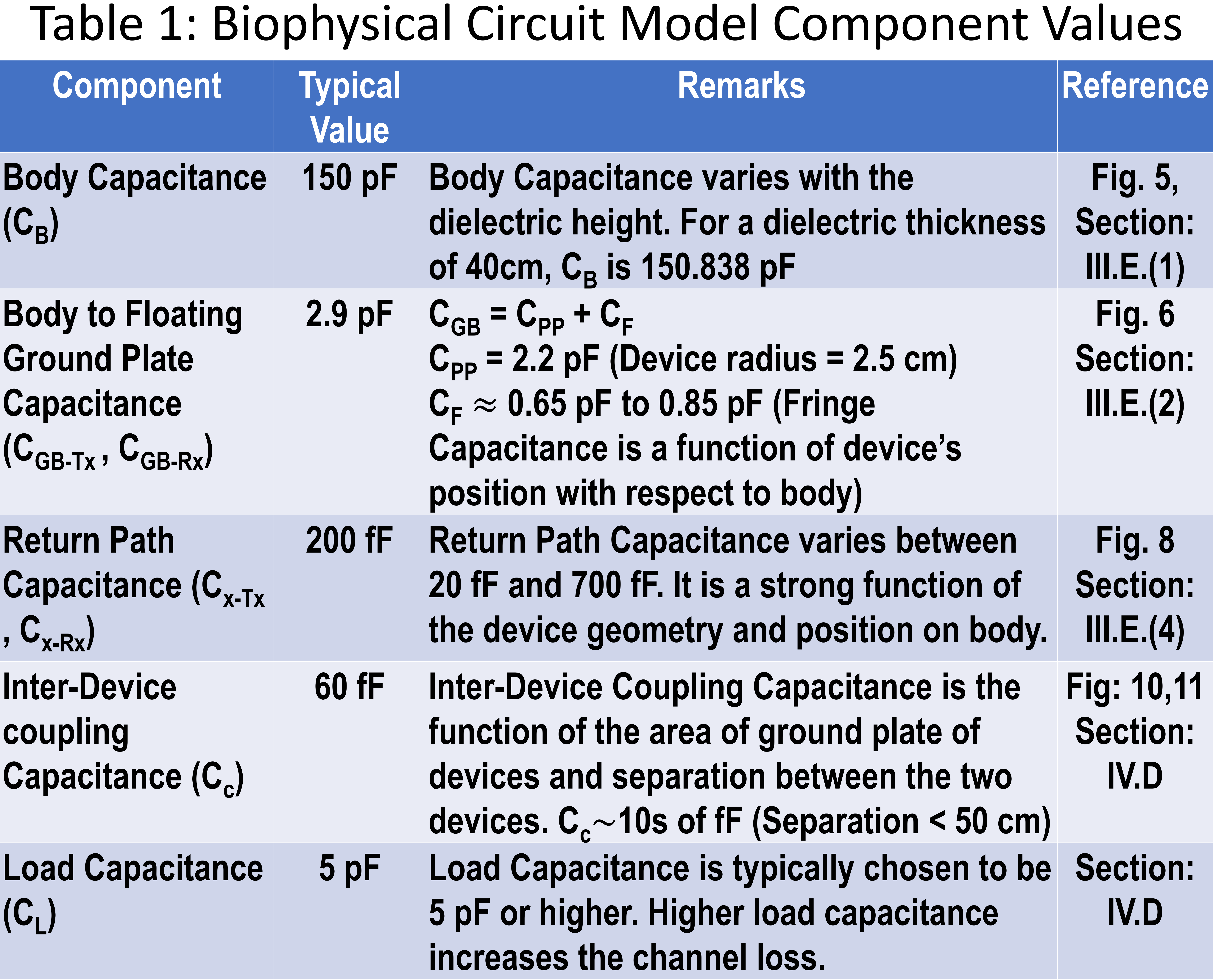}
\label{fig:table_components}
\end{figure}

\subsection{Path Loss as a Function of Device Geometry}

The path loss for a capacitive EQS HBC system is provided by Eqn: \ref{eqn:Inter_Device_Coupling}. This equation can be related to the geometry and the position of the transmitting and receiving devices. Return path capacitance as a function of the device dimension is provided by Eqn: \ref{Eqn:return_path_cap}. Further, Eqn: \ref{Eqn:Coupling_Cap} shows how the position and dimension of device changes the coupling capacitance. The plate-to-plate capacitance as a function of the device geometry is shown by Eqn: \ref{eqn:Cpp}.\par
Considering the transmitter and receiver to be of the same radius, the voltage transfer function can be represented as:

\begin{equation}
    \frac{V_{o}}{V_{in}} = \frac{k\frac{\pi a^2}{d}+ \frac{x_{Tx}x_{Rx}(8\epsilon_0 a)^2}{C_B}}{k\frac{\pi a^2}{d} + \frac{\epsilon_0 \pi a^2}{t} + C_F + C_L}
    \label{eqn:Inter_Device_Coupling_geometry}
\end{equation}

When the devices are placed away from each other, the voltage transfer function equation can be simplified as shown in Eqn: \ref{eqn:inter_device_coupling_simplified}. This can be further represented as:
\begin{equation}
    \frac{V_{o}}{V_{in}} = \frac{ x_{Tx}x_{Rx}(8\epsilon_0 a)^2}{C_B (\frac{\epsilon_0 \pi a^2}{t} + C_F + C_L)}
    \label{eqn:Inter_Device_Coupling_geometry_simplified}
\end{equation}

Eqn: \ref{eqn:Inter_Device_Coupling_geometry} and \ref{eqn:Inter_Device_Coupling_geometry_simplified} describe the channel loss of a capacitive HBC working in the Electro Quasistatic (EQS) domain as a function of the size of the devices (radius - $a$), the distance between the two devices ($d$), the thickness of the receiver ($t$) as well as the position of the devices ($x_{Tx}$, $x_{Rx}$) with respect to the human body.

\section{Experimental Results}

\subsection{Experimental Setup}
\begin{figure}[h!]
\centering
\includegraphics[width=0.45\textwidth]{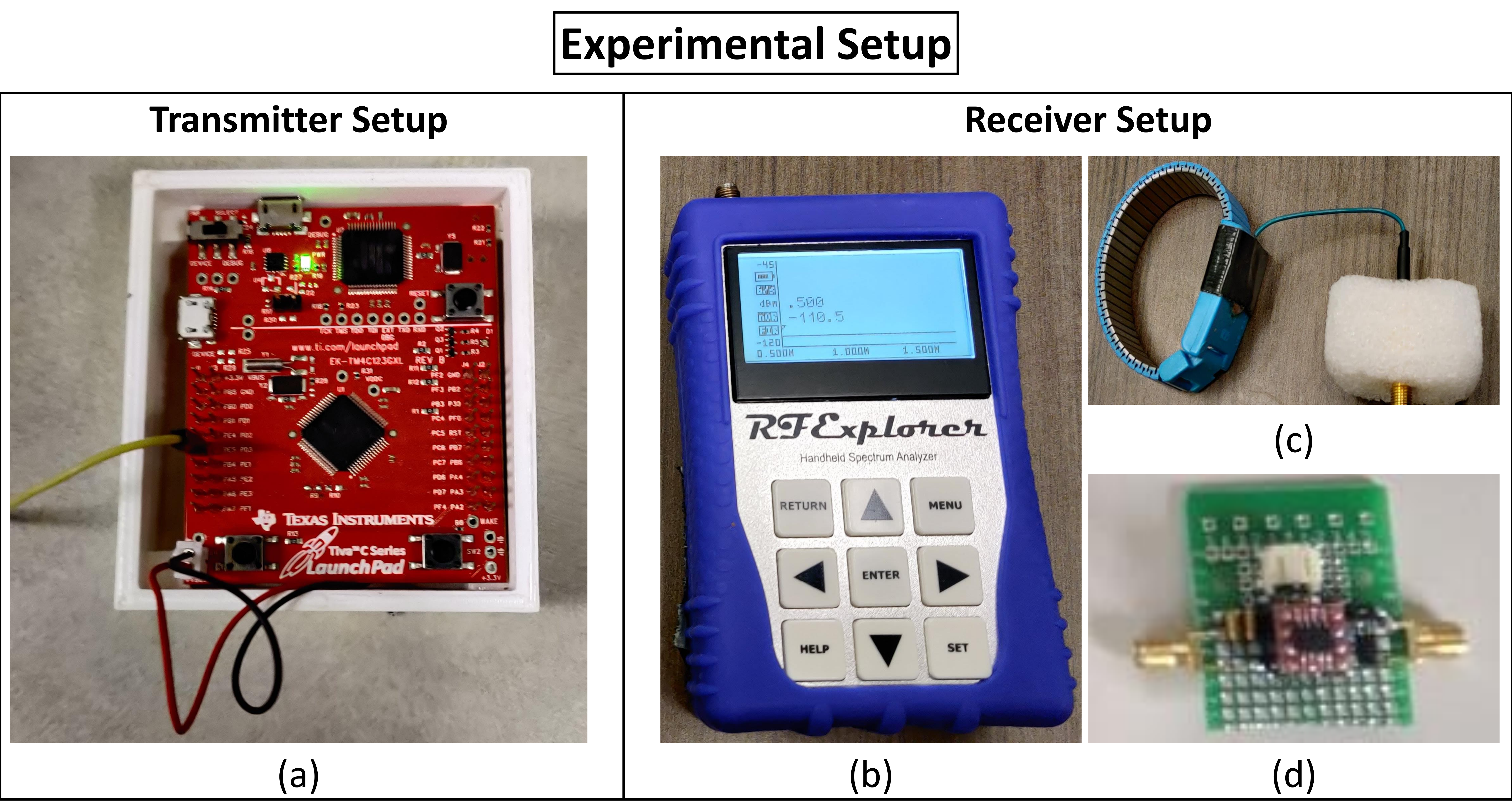}
\caption{Transmitter Setup: (a) Texas Instruments Tiva C board contained in a 3D printed case for low frequency (1MHz) signal transmission. Receiver Setup: (b) handheld RF Explorer spectrum analyzer to capture the received signal, (c) couplers used to connect the receiver to the body, (d) buffer developed in-house is connected to the receiver to provide a high load impedance.}
\label{fig:Expts_Setup}
\end{figure}

The experiments were carried out in a lab environment. This study was approved by the Institute Review Board (IRB). The transmitting device used is a signal generator developed in-house using a Texas Instruments Tiva C Launchpad evaluation board which is enclosed in a 3D printed case as shown in Fig. \ref{fig:Expts_Setup} (a). The operating frequency of the low frequency signal generator is at 1 MHz.\newline

The receiver used for the measurements is a handheld RF Explorer spectrum analyzer (Fig. \ref{fig:Expts_Setup} (b)). The couplers (Fig. \ref{fig:Expts_Setup} (c)) shown are used to connect the spectrum analyzer to the body. A high frequency buffer (Fig. \ref{fig:Expts_Setup} (d)) developed in-house is used to get a high load impedance at the receiver end. The ground plate of the receiver device is the outer body of the spectrum analyzer which has an area of about $80 cm^2$. The simulations shown are performed with devices of much smaller sizes with an area of less than $30 cm^2$. This increased size of ground plate would increase the received voltage as predicted by Eqn: \ref{eqn:Inter_Device_Coupling_geometry} which relates the path loss of a capacitive HBC channel in terms of the sizes of the devices used.

\begin{figure*}[t!]
\centering
\includegraphics[width=\textwidth]{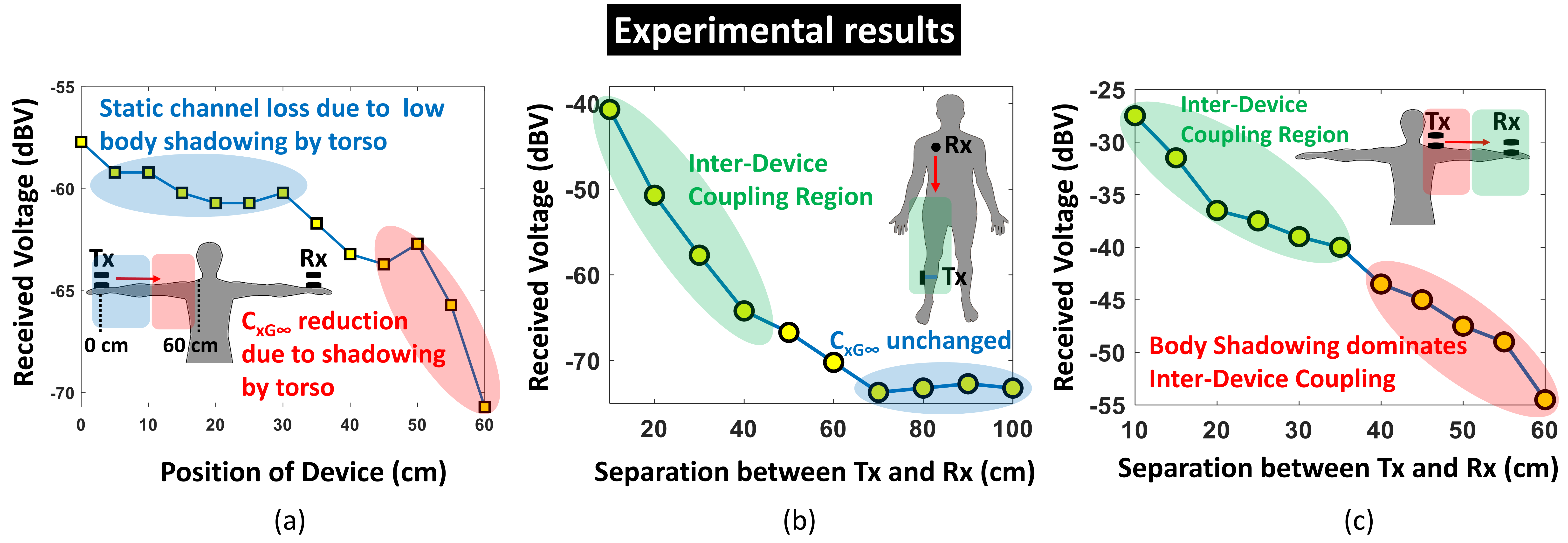}
\caption{(a) Transmitter and receiver is placed on opposite arms to nullify the effects of inter-device coupling. Body shadowing effects is observed when the transmitter is placed close to the torso. (b) Receiver and transmitter is placed on the torso while the separation between them is reduced. Body shadowing by torso is constant across all positions of the devices showing a static channel loss when the devices are placed away from each other. However, inter-device coupling dominates when the devices are in close proximity decreasing the channel loss. (c) A combination of shadowing by torso and inter-device coupling is observed when the transmitter and receiver are placed on the same arm. When the separation is large, body shadowing dominates inter-device coupling causing channel loss to decrease. When the separation is reduced, inter-device coupling starts dominating resulting in increasing channel loss.}
\label{fig:Expt_Results}
\end{figure*}

\subsection{Body Shadowing Effects}
As observed from simulations shown in Fig. \ref{fig:Rx_Circuit} (c), the received voltage on the body decreases when we bring the transmitter or the receiver close to the torso. This has been concurrent with our experimental findings as shown in Fig. \ref{fig:Expt_Results} (a) where we see that there is a sudden drop in the received voltage due to reduction in the return path capacitance because of body shadowing effects.

The received voltage is measured using the spectrum analyzer which has a bigger ground plate which results in a lower channel loss with respect to the simulations. We also observe a flat band response when the transmitter is placed away from the torso. This is because the effect of the arm in blocking the field lines originating from the ground plate of the transmitter is constant when the device is away from the torso. It has also been observed that the channel loss is lower for the corner case where the transmitter is placed at the end of the arm as the device is now more exposed to the earth's ground.

\subsection{Inter-Device Coupling}

The effects of inter-device coupling and body shadowing is observed experimentally in two different situations as shown in Fig. \ref{fig:Expt_Results} (b) and (c).

In the first case shown in Fig. \ref{fig:Expt_Results} (b), the transmitter and receiver are placed on the torso. Here, the transmitter is tied to the leg while the receiver is moved across the torso thereby decreasing the separation between the devices. We see that initially when the two devices are placed at a separation of $100 cm$ from each other, we observe no effects of inter-device coupling. It is also observed that the received voltage on the torso doesn't vary to a great degree when the two devices are placed far apart showing that the return path capacitance is constant on the torso. Inter-device coupling capacitance starts affecting the value of the channel loss as the separation between the two devices is decreased to lower than $50 cm$.  \par
Fig. \ref{fig:Expt_Results} (c) illustrates the scenario where the transmitter and receiver devices are placed on the same arm. When the transmitter is placed close to the body, the effects of body shadowing dominates the weak inter-device coupling. A weak inter-device coupling is present between the devices as the separation between the transmitter and receiver is small and there is a direct line-of-sight between the two devices. When the transmitter is brought closer to the receiver, inter-device coupling increases as the separation between devices decreases. When the transmitter is moved away from the torso, inter-device coupling starts dominating due to no body shadowing effect from the torso. In this scenario, the channel loss decreases rapidly due to a capacitive path between the two devices.  

 
\begin{figure}[h!]
\centering
\includegraphics[width=\columnwidth]{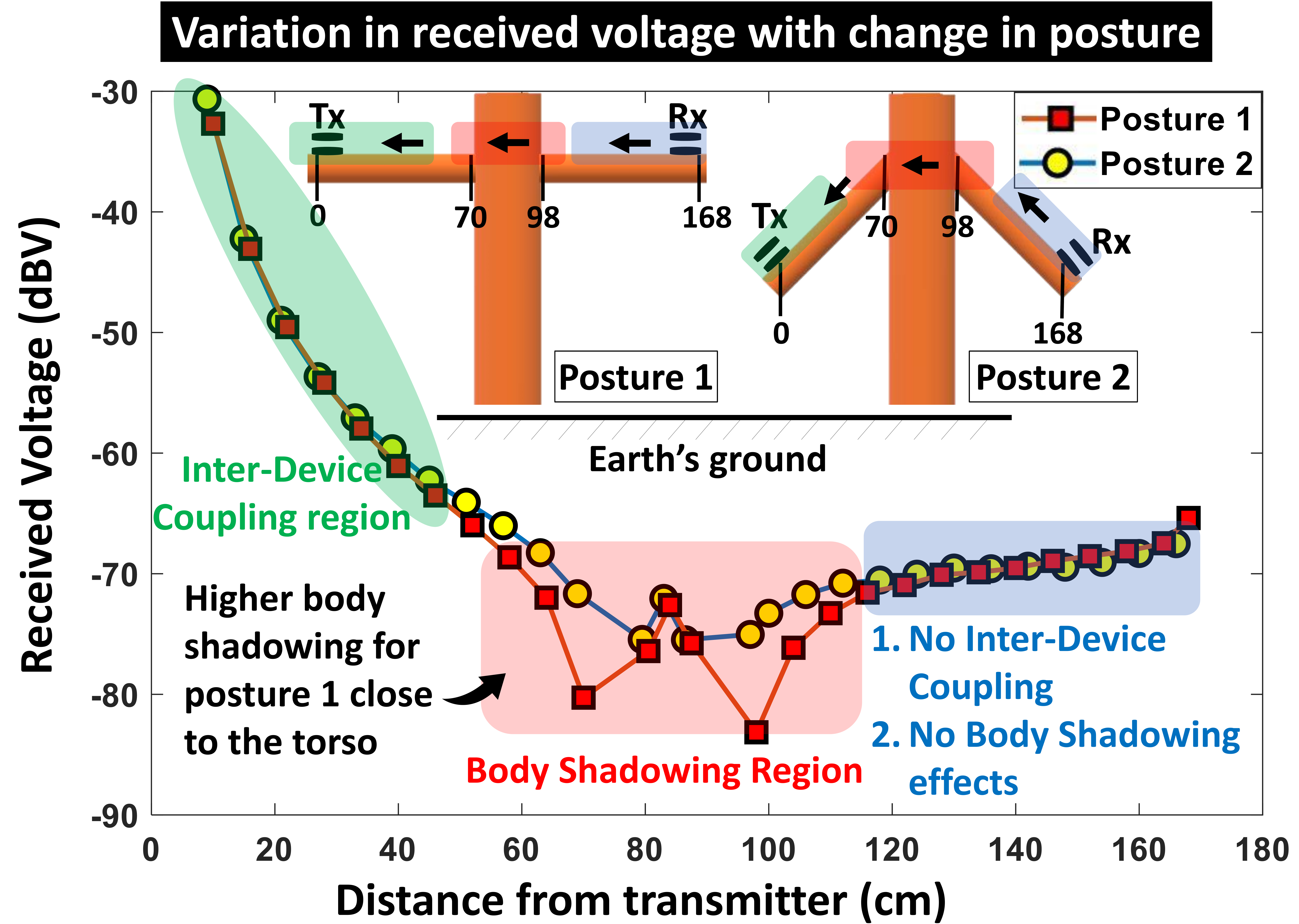}
\caption{Variability in channel loss with change in position of receiver is illustrated for two different postures. Posture 1 has higher loss than posture 2 around the shoulder region close to the torso due to higher body shadowing in the model shown. The variation in postures doesn't significantly change the channel loss when the receiver is away from the torso or on the torso. Further, the channel loss variation due to inter-device coupling can be observed to be much larger than due to body shadowing.}
\label{fig:Posture_Variation}
\end{figure}
\section{Conclusion}

We present an advanced biophysical model which captures the variability in channel characteristics for a capacitive EQS HBC system. The channel loss which is a strong function of the return path capacitance is analyzed in detail. The change in return path capacitance is attributed to the shadowing effects of the human body. The effects of inter-device coupling when two devices are close to each other is also investigated.\par
This understanding of body shadowing and inter-device coupling can be used to explain the channel loss characteristics observed for different postures intuitively as shown in Fig. \ref{fig:Posture_Variation}. The region shaded red shows body shadowing affecting posture $1$ more than posture $2$.  In posture $2$, the arms are bent at an angle towards the earth's ground plane. When the receiver is placed in the part of the model shaded red (shoulder region), the floating ground plate of the receiver faces away from the torso tilting towards the earth's ground plane thus reducing the effect of body shadowing thus reducing channel loss in posture 2. However, when the receiver is on the torso, we see no significant difference in the channel loss for the two cases as body shadowing on torso is equal for both. Similarly, when the two devices are close to each other (region shaded green), inter-device coupling dominates body shadowing and affects both the postures equally thus resulting in no significant change in channel loss. When the receiver is away from the torso and on the opposite arm to where the transmitter is placed (region shaded blue), channel loss is same for both the postures as both body shadowing due to torso and inter-device coupling are ineffective in this region. The curve also shows the extent of channel loss variability due to inter-device coupling and body shadowing effects. Inter-device coupling can be observed to have a far greater impact on channel variability than body shadowing effects.

 \par
Further, the updated biophysical model is used to find out a closed form equation of the path loss for a capacitive HBC channel in the EQS domain. The parameters used in the biophysical model are then found out as a function of the device geometry as well as the position of the devices with respect to the human body which can be used to design and develop wearable WBAN systems using capacitive electro-quasistatic human body communication.

\section*{Acknowledgment}

This work was supported in part by the Air Force Office of Scientific Research YIP Award under Grant FA9550-17-1-0450 and the National Science Foundation Career Award under Grant CCSS 1944602. We would like to thank Dr. Shovan Maity, graduated PhD student, Baibhab Chatterjee, Nirmoy Modak, Debayan Das and Shitij Avlani, PhD students at Purdue University for their immense cooperation and support during the experiments.

\ifCLASSOPTIONcaptionsoff
  \newpage
\fi



%

\bibliographystyle{IEEEtran}
\bibliography{IEEEabrv,references.bib}

\end{document}